\documentclass[aps,pra, reprint, superscriptaddress, showpacs]{revtex4-1}

\usepackage{graphicx}
\usepackage{amssymb}
\usepackage{epstopdf}
\usepackage{amsmath}

\begin{document}

\title{Upper Bound to the Ionization Energy of $^{85}$Rb$_2$}
\author{M. A. Bellos}
\author{R. Carollo}
\author{J. Banerjee}
\author{M. Ascoli}
\affiliation{Department of Physics, University of Connecticut, Storrs, Connecticut 06269, USA}
\author{A.-R. Allouche}
\affiliation{Institut Lumi\`{e}re Mati\`{e}re, UMR5306 Universit\'{e} Lyon 1-CNRS, Universit\'{e} de Lyon 69622 Villeurbanne cedex, France}
\author {E. E. Eyler}
\author{P. L. Gould}
\author{W. C. Stwalley}
\affiliation{Department of Physics, University of Connecticut, Storrs, Connecticut 06269, USA}

\date{\today}

\begin{abstract}
We report an upper bound to the ionization energy of $^{85}$Rb$_2$ of 31$\,$348.0(6) cm$^{-1}$, which also provides a lower bound to the dissociation energy $D_0$ of $^{85}$Rb$_2^+$ of 6$\,$307.5(6) cm$^{-1}$. These bounds were measured by the onset of autoionization of excited states of $^{85}$Rb$_2$ below the 5$s$+7$p$ atomic limit. We form $^{85}$Rb$_2$ molecules via photoassociation of ultracold $^{85}$Rb atoms, and subsequently excite the molecules by single-photon ultraviolet transitions to states above the ionization threshold.
\end{abstract}


\maketitle

\section{Introduction}

\begin{figure*}[]
\includegraphics[scale=0.8]{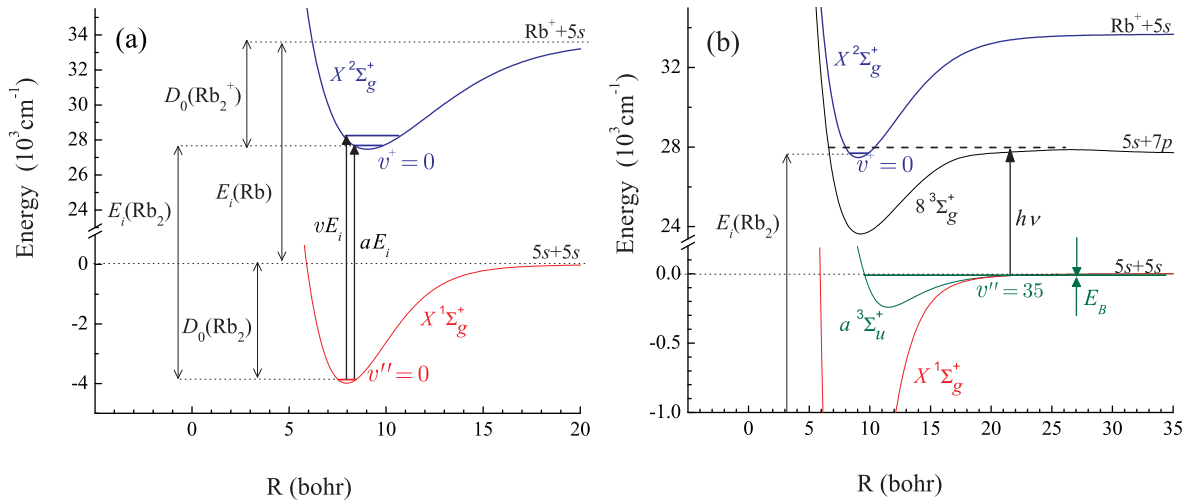}
\caption{\label{pecs} (Color Online) (a) Definition of various ionization and dissociation energies. $E_i$(Rb$_2$) is the energy interval between the ground state of the molecule ($X\,^1\Sigma_g^+$, $v''$=$J''$=$0$) and the ground state of the molecular ion ($X\,^2\Sigma_g^+$, $v^+$=$N^+$=$0$). $vE_i$ and $aE_i$ are the vertical ionization energy and adiabatic ionization energy, respectively. (b) Relevant potential energy curves (from \cite{strauss10} and this work). In our measurement the initial state for photoexcitation is $a\,^3\Sigma_u^+$, $v''$ = 35 . The dashed line indicates the energy region of the autoionizing levels observed in our spectra.}
\end{figure*}

The ionization energy, also known as the ionization potential, is the minimum energy required to ionize a ground-state atom or molecule. In the case of molecules, one can distinguish between two kinds of ionization energy: the `adiabatic' ionization energy ($aE_i$) and the `vertical' ionization energy ($vE_i$). The $aE_i$ is the energy required to reach the lowest ionization threshold, i.e. the rovibrational ground state of the ion, as shown in Fig. \ref{pecs}(a). The $vE_i$ is the lowest \emph{observed} ionization energy and often correlates to a rovibrationally excited ionization threshold. The $vE_i$ is greater than or equal to the $aE_i$, and depends on the Franck-Condon factor for the ionizing transition and the limit of experimental signal-to-noise. The $aE_i$, therefore, is a more fundamental quantity, and so we will refer to it here as the ionization energy ($E_i$).\\

The ionization energy of Rb$_2$ is currently not well known. Although measurements \cite{lee65, klucharev80, wagner85, kappes85} and calculations \cite{preuss55, bellomonte74,patil00,silberbach86,szentpaly82,krauss90,aymar03, jraij03} exist, they have large uncertainties. This is in contrast to other alkali dimers, e.g., Li$_2$ \cite{roche93}, Na$_2$ \cite{chang99}, K$_2$ \cite{broyer83}, and Cs$_2$ \cite{kim93}, where the $E_i$'s have been measured to accuracies between 0.02 and 2 cm$^{-1}$ (see Ref. \cite{stwalley93} for a review).
In an effort to measure the $E_i$ of Rb$_2$, we have performed spectroscopy of ultracold Rb$_2$ in the energy region corresponding to the bottom of the ground-state potential well of the Rb$_2^+$ molecular ion. This measurement also allows us to search for efficient pathways for the production of ultracold Rb$_2^+$ in selected rovibrational levels of its ground state (e.g. $v^+$=$N^+$=$0$). Part of the appeal of working with molecular ions is that they share many of the features of neutral molecules and have the added benefit of long trapping lifetimes in ion traps.\\

There are several methods to measure $E_i$'s, such as the extrapolation of a Rydberg series or the observation of the onset of direct photoionization or autoionization (Ref. \cite{broyer83} contains examples of each of these methods). These methods rely on two distinct ionization mechanisms, direct photoionization and autoionization \cite{[{}] [{, Chap. 7.}]lefevbre-brion04}. Direct photoionization, or photoionization for short, proceeds through a single step,

\begin{equation}\label{photoionization}
\mathrm{Rb}_2 + h\nu \rightarrow \mathrm{Rb}_2^+ + e^-,
\end{equation}
whereas autoionization,

\begin{equation}\label{photoexcitation}
\mathrm{Rb}_2 + h\nu \rightarrow \mathrm{Rb}_2^{**} \rightarrow \mathrm{Rb}_2^+ + e^-,
\end{equation}
proceeds through an intermediate state that spontaneously ionizes. Here the notation Rb$_2^{**}$ denotes ``superexcited'' levels of the molecule above the ionization threshold as shown in Fig. \ref{pecs}(b). Transition (1) is bound-free and generally shows broad continuum features, while transition (2) includes a bound-bound step and can show sharp features. Neither transition is possible unless the photon energy is above the ionization energy.\\

The $E_i$ of Rb$_2$ is related to the dissociation energy of its molecular ion, $D_0(\mathrm{Rb}_2^+)$, the dissociation energy of the neutral molecule, $D_0(\mathrm{Rb}_2)$, and the $E_i$ of the atom, $E_i$(Rb), via the relationship,
\begin{equation}\label{iegeneral}
E_i\mathrm{(Rb}_2)+D_0(\mathrm{Rb}_2^+)=E_i\mathrm{(Rb)}+D_0(\mathrm{Rb}_2),
\end{equation}
as is shown graphically in Fig \ref{pecs}(a).

We use a variation of Eq. (\ref{iegeneral}) to account for the fact that we photoexcite from an excited state instead of the absolute ground state of Rb$_2$ ($X\,^1\Sigma_g^+$, $v''$=$J''$=0). Therefore,

\begin{equation}\label{ieparticular}
E_i\mathrm{(Rb}_2)=  h\nu +E_B + D_0(\mathrm{Rb}_2),
\end{equation}
where $h\nu$ is the energy of the applied photon and $E_B$ is the (negative) binding energy of the initial state, defined with respect to the 5$s$+5$s$ atomic limit. The energy of the observed superexcited state corresponds to $h\nu +E_B$. By adding the dissociation energy $D_0(\mathrm{Rb}_2)$, we shift the energy reference from the atomic limit to the $X\,^1\Sigma_g^+$ ($v''$=$J''$=0) level.

We are not aware of any direct measurements of $D_0(\mathrm{Rb}_2)$. We can however calculate accurate values for $D_0(\mathrm{Rb}_2)$ and $E_B$ using the \emph{LEVEL8.0} program \cite{level8} and potential energy curves based on fits to numerous spectroscopic measurements \cite{amiot93,tsai97,seto00,beser09,strauss10}. The most recent work in this series of spectroscopic measurements and fits, by Strauss $et$ $al.$ \cite{strauss10}, reports accuracies of 50 MHz for deeply-bound levels and a few MHz for levels near the dissociation limit.\\

\begin{figure*}[]
\includegraphics[scale=0.8]{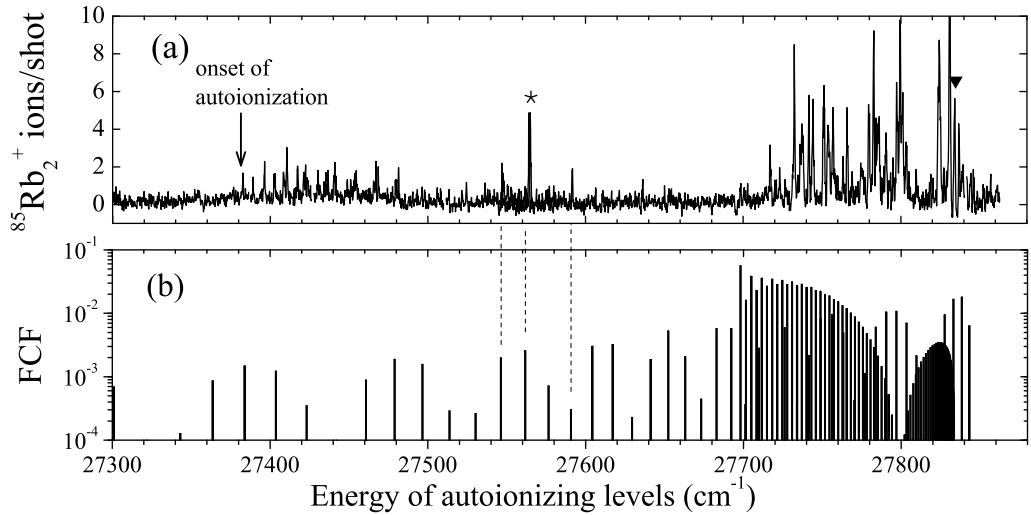}
\caption{\label{ipspectrum} (a) Autoionizing levels of $^{85}$Rb$_2$ photoexcited from the $v''$ = 35 level of the $a\,^3\Sigma_u^+$ state. The horizontal axis is the sum of the photon energy and the binding energy of the initial level, which gives the energy of the autoionizing levels above the 5$s$+5$s$ atomic limit. The arrow (\textbf{$\downarrow$}) shows the lowest-energy line reproducible via autoionization. The star ($\star$) marks the line used to study power dependence. The triangle ($\blacktriangledown$) labels a line originating from two-photon ionization of Rb through the $\left |7p_{1/2} \right>$ state. This atomic transition is strong enough to create a spurious signal in the Rb$_2^+$ time-of-flight window, and marks the position of the 5$s$+7$p_{1/2}$ atomic limit. (b) Simulated spectrum generated by plotting the Franck-Condon factors (FCFs) between the initial $v''$ = 35 level and various vibrational levels of the excited state, as a function of the excited-level energy. We have shifted the simulated spectrum energy to match the position of the 5$s$+7$p_{1/2}$ atomic limit. The FCFs and vibrational level energies were calculated using the \emph{LEVEL8.0} program \cite{level8}. Correlations between the observed and simulated spectra are discussed in the text.}
\end{figure*}

\section{Experiment}

The apparatus used for this measurement has been described previously \cite{bellos12} and is only briefly summarized here. The starting point is a magneto-optical trap (MOT) that traps about $8\times10^7$ atoms at a peak density of $1\times10^{11}$ cm$^{-3}$ and a temperature of 120 $\mu$K. The MOT is continuously irradiated by a photoassociation (PA) laser to convert a fraction of the trapped atoms into molecules. After we photoassociate atoms into excited-state molecules, they decay radiatively and populate the metastable $a\,^3\Sigma_u^+$ state. We form molecules in specific vibrational levels of the $a\,^3\Sigma_u^+$ state via PA through the 1($0_g^-$)$\,$($v'\simeq$ 173, $J'$=1) level, detuned 17.1 cm$^{-1}$ below the 5$s$+5$p_{1/2}$ atomic limit \cite{lozeille06}. This results in the formation of $a\,^3\Sigma_u^+$ state molecules primarily in the ($v''$ = $35$, $J''$ = $0$) and ($v''$ = $35$, $J''$ = $2$) levels, bound by $-0.806(2)$ and $-0.794(2)$ cm$^{-1}$, respectively. Averaging the energy over the two rotational levels yields a binding energy of $E_{B}$ = $-0.800(6)$ cm$^{-1}$ for $v''$ = $35$ molecules. The distribution of vibrational levels was measured to be 70$\%$ in $v''$ = $35$, and around 10$\%$ each in the neighboring vibrational levels $v''$ = 34, 36, and 37. This distribution was determined by fitting lineshapes to the REMPI spectra of Ref. \cite{lozeille06}. It is important to measure this distribution, rather than using calculated Franck-Condon factors for radiative decay, because the PA laser strongly modifies the distribution of the upper-most vibrational levels \cite{huang06}.

Although molecules are continuously produced in the MOT by the PA laser, they are also continuously lost because they are not well trapped by the MOT. We periodically photoexcite the molecules that remain with a pulsed ultraviolet laser. The steady-state number of molecules within the $\sim\,$4$\,$mm diameter uv laser beam is approximately 100. This small number of molecules is sufficient due to the high quantum efficiency of ion detection. The uv light is tuned around 365 nm and is produced by frequency doubling an infrared pulsed dye laser. The pulsed dye laser is operated using a LDS750 dye solution and pumped by a doubled Nd:YAG laser (532 nm, 10 ns pulses at 10 Hz repetition rate). A frequency doubler (Inrad Autotracker III) produces the second harmonic of the pulsed dye laser with roughly 25\% efficiency, yielding a uv pulse energy $\sim$ 1 mJ/pulse. The measured uv pulse linewidth is 0.9 cm$^{-1}$, about twice that of fundamental infrared pulse.

After the atoms and molecules are ionized, they travel to an ion detector where Rb$^+$ and Rb$_2^+$ ions are distinguished by their time of flight. A boxcar integrator monitors the arrival of Rb$_2^+$ ions 15 $\mu$s after the uv pulse. We switch off the MOT lasers 20 $\mu$s before the arrival of the uv pulse, so as to depopulate the atomic $\left |5p_{3/2} \right>$ state and suppress the production of Rb$^+$ ions. Any Rb$^+$ thus produced originates from two-photon off-resonant ionization. Rb$_2^+$ is produced by one-photon processes as discussed below.\\

\begin{figure}[]
\includegraphics[scale=0.8]{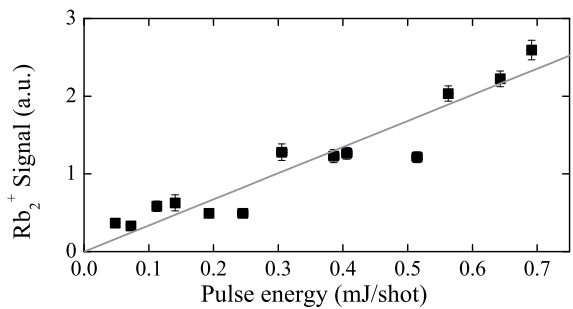}
\caption{\label{onephoton}   Power dependence of a selected spectral line (marked by a star ($\star$) in Fig. \ref{ipspectrum}(a)), along with a straight line fit.}
\end{figure}

\section{Autoionization spectroscopy and Results}

By scanning the uv laser and monitoring the production of Rb$_2^+$, we obtain the spectrum shown in Fig. \ref{ipspectrum} (a). The laser power dependence of the signal is linear, as shown in Fig. \ref{onephoton}, comfirming that a one-photon transition is responsible for the ionization. Furthermore, the absorption of a second photon would further excite the molecules to an energy region dominated by repulsive curves of electronically-excited Rb$_2^+$ \cite{jraij03}. This would be expected to dissociate the molecule without producing a Rb$_2^+$ signal. By ruling out two-photon transitions, we confirm that the observed lines are above the one-photon ionization threshold.\\

There is no evidence for direct photoionization in the spectrum, as there is no broad Rb$_2^+$ background signal or continuum threshold. The lack of photoionization indicates that high-$n$ Rydberg states are probably not populated, because of the continuity of oscillator strength across the ionization threshold. More specifically, the oscillator strength per unit energy to the continuum just above the ionization threshold is equal to the oscillator strength to high-$n$ Rydberg states just below the threshold \cite{[{}] [{, pp. 40-45.}]gallagher94}. A further indication that Rydberg states (high-$n$ or low-$n$) are not being observed is that the spacing between lines is nearly regular and does not correspond to the $1/n^3$ spacing of a Rydberg series. The absence of photoionization suggests that higher molecule numbers will be necessary to accurately measure the ionization energy through the onset of photoionization or Rydberg series extrapolation. Also, the absence of photoionization allows us to place an upper bound to the photoionization cross-section of the initial state ($\sigma$ $<$ $5\times10^{-19}$ cm$^{2}$). Although photoionization and autoionization can simultaneously occur above the ionization threshold, autoionization has been more prevalent than photoionization in ultracold experiments to date (see, for example, Refs. \cite{machholm94, amelink00, trachy07}).\\

In an effort to assign the observed spectrum, we calculated \textit{ab initio} potential energy curves (PECs) leading to the 5$s$+7$p$ atomic limit using the method of Allouche and Aubert-Fr\'{e}con \cite{allouche12}. We extended the basis set used in Ref. \cite{allouche12} by adding one $f$-orbital, with exponent 0.1, and set the cutoff parameter of the core polarization potential for the $f$-orbital to 2.5025 $a_0$. We include the eight PECs ($^1\Sigma_u^+$, $^3\Sigma_u^+$, $^1\Pi_u$, $^3\Pi_u$, $^1\Sigma_g^+$, $^3\Sigma_g^+$, $^1\Pi_g$, $^3\Pi_g$) that correlate to the 5$s$+7$p$ atomic limit, and the Rb$_2^+$ ground state PEC also calculated for this work, as supplementary material to this paper.\\

We can rule out transitions to the four $ungerade$ excited states by applying the $u\leftrightarrow g$ electric dipole selection rule. In Fig. \ref{ipspectrum}(b) we plot a simulated spectrum generated from the PECs of the initial $a\,^3\Sigma_u^+$ state and the excited $^3\Sigma_g^+$ state. Although transitions to the other three $gerade$ excited states are, in principle, allowed, the observed spectrum does not correlate well with simulated spectra from these PECs. The simulated spectrum to the $^3\Sigma_g^+$ state reproduces three features of the observed spectrum: (1) the large line spacings between $24\,400$ and $27\,700$ cm$^{-1}$ as shown by the dashed vertical lines, (2) the high density of lines between $27\,700$ cm$^{-1}$ and the atomic limit at $27\,835$ cm$^{-1}$, (3) the presence of a quasibound level just above the atomic limit.

Lines in Fig. \ref{ipspectrum}(b) below $27\,700$ cm$^{-1}$ correspond to the inner well of the excited state, while lines above $27\,700$ cm$^{-1}$ correspond to both the inner and outer wells of the excited state. The closely spaced lines between $27\,700$ and $27\,800$ cm$^{-1}$ correspond to transitions from the outermost lobe of the initial state wavefunction (at the outer turning point). The closely spaced lines between $27\,800$ and $27\,835$ cm$^{-1}$ correspond to transitions from the second-to-last lobe of the initial state wavefunction.  This simulated spectrum does not include effects such as a $R$-dependent transition dipole moment, spin-orbit coupling effects, tunneling between wells, or avoided crossings between PECs. Further analysis of this spectrum will be the subject of future work.\\

\begin{figure}[]
\includegraphics[scale=0.40]{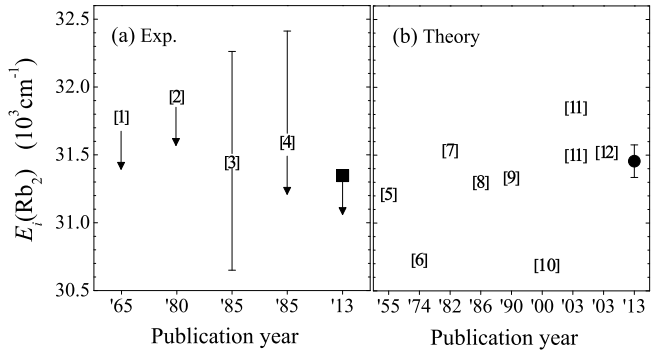}
\caption{\label{experimentalips} Experimental and theoretical ionization energies of Rb$_2$. (a) Experimental measurements with the present measurement labeled by the square ({\tiny$\blacksquare$}). Refs. \cite{lee65,kappes85} are measurements of vertical ionization energies, and hence upper bounds to the ionization energy. Ref. \cite{klucharev80} describes a measurement of the ionization energy, but is characterized in Ref. \cite{stwalley93} as a vertical ionization energy instead. (b) Theoretical calculations with the present calculation labeled by the bullet ($\bullet$). Refs. \cite{wagner85,preuss55,bellomonte74,patil00,silberbach86,szentpaly82,krauss90,aymar03, jraij03} report dissociation energies of Rb$_2^+$, which we convert to ionization energies (see text). The two values reported by Ref. \cite{aymar03} correspond to different approximations used.}
\end{figure}

Regardless of the spectral assignment, we can use the spectral line with the lowest observed energy to place an upper bound on the $E_i$. This line, identified by an arrow in Fig. \ref{ipspectrum}(a), corresponds to a photon of energy 27$\,$384.0(3) cm$^{-1}$ exciting a molecule bound by $E_B$ = $-0.8$(5) cm$^{-1}$. We have increased the uncertainty in the binding energy of the initial level from $\pm\,$0.006 to $\pm\,$0.5 cm$^{-1}$, to account for the small possibility that the signal may originate from vibrational levels adjacent to $v''$ = 35. These adjacent levels are populated in small quantities as discussed in Sec. II. We use this line energy and a calculated value of $D_0(^{85}\mathrm{Rb}_2)$ = 3$\,$964.74(2) cm$^{-1}$ derived from Ref. \cite{strauss10} in Eq. (\ref{ieparticular}) to set an upper bound to $E_i$($^{85}$Rb$_2$) of 31$\,$348.0(6) cm$^{-1}$. This upper bound is more constraining than previous measurements \cite{lee65, klucharev80, wagner85, kappes85} and is plotted in Fig. \ref{experimentalips}(a). With the same line energy and $E_i$($^{85}$Rb) = 33$\,$690.797$\,$5(2) cm$^{-1}$ \cite{sanguinetti09}, we can set a lower bound to $D_0$($^{85}$Rb$_2^+)$ of $E_i$($^{85}$Rb) $-$ ($h\nu+E_B$) = 6$\,$307.5(6) cm$^{-1}$.

The presently calculated Rb$_2^+$ ground state PEC has a theoretical dissociation energy $D_0(^{85}\mathrm{Rb}_2^+)$ = 6200 cm$^{-1}$, computed using \emph{LEVEL8.0}. It is difficult to accurately know the uncertainty for this value. Nevertheless we can \emph{estimate} the theoretical uncertainty by comparing differences between theoretical and available experimental dissociation energies as was done in Ref. \cite{allouche12}. Doing this, we find an average error in the dissociation energy of 1.9 $\%$ of the well depth, corresponding to $\pm$ 120 cm$^{-1}$ for $D_0(^{85}\mathrm{Rb}_2^+)$ = 6200 cm$^{-1}$. Using Eq. (\ref{iegeneral}) we can easily convert this dissociation energy into an ionization energy, with negligible increases in uncertainty, as $D_0(^{85}\mathrm{Rb}_2)$ and $E_i$($^{85}$Rb) are known to within 60 MHz and 6 MHz, respectively. This yields a theoretical $E_i$($^{85}$Rb$_2$) of 31$\,$456 $\pm$ 120 cm$^{-1}$ which we plot in Fig. \ref{experimentalips}(b) alongside with previous theoretical values \cite{preuss55,bellomonte74,patil00,silberbach86,szentpaly82,krauss90,aymar03, jraij03}.\\

If we assume that the theoretical $E_i$ values and associated uncertainties are accurate, we can alternatively use the observed onset of autoionization to determine which vibrational levels are populated in the molecular ion. Our theoretical lower bound of $E_i$, 31$\,$336 cm$^{-1}$, is below the observed onset by only 12 cm$^{-1}$. This difference is smaller than the vibrational spacing of 46 cm$^{-1}$ for the first few vibrational levels, implying that the produced ions are possibly in the $v^+$ = 0 level.

We expect these ions to be slightly hotter than the atoms in the MOT, due to the energy released when Rb$_2^{**}$ autoionizes into Rb$_2^+$. This heating should, in principle, not significantly reduce the trapping lifetime for deeply-trapped ions, and can be minimized by ionizing as close to the threshold as possible.\\

It should be noted that we have also photoexcited $a\,^3\Sigma_u^+$ ($v''$ = 0) molecules to the same spectral region. We produced $a\,^3\Sigma_u^+$ ($v''$ = 0) molecules via blue-detuned photoassociation at short internuclear distances \cite{bellos11}. In the case of photoexcitation starting from $v''$ = 0, we observed an onset of autoionization 98.1 cm$^{-1}$ higher than that observed when starting from $v''$ = 35. The measurement starting from $v''$ = 0, therefore, provides a less constraining bound, despite the fact that photoexcitation occurs at shorter internuclear distances.\\

\section{Conclusion}
We report an improved upper bound to the ionization energy of $^{85}$Rb$_2$, $E_i$($^{85}$Rb$_2$) $ \leq$ 31$\,$348.0(6) cm$^{-1}$ and a corresponding lower bound to the dissociation energy of the molecular ion $^{85}$Rb$_2^+$, $D_0$($^{85}$Rb$_2^+)$ $\geq$ $6$\,$307.5(6)$ cm$^{-1}$. Measuring the $E_i$ directly rather than setting an upper limit will require a measurable photoionization signal at threshold, or alternatively a well resolved series of Rydberg states. Such a signal may become observable by replicating the experiment in an optical trap, where the number and density of molecules are orders of magnitude greater than in the present experiment.\\

\begin{acknowledgments}
We would like to thank D. Rahmlow and L. Aldridge for laboratory assistance and gratefully acknowledge funding from NSF (Grants No. PHY-0855613 and PHY-1208317), AFOSR MURI (Grant No. FA 9550–-09–-1–-0588), and the University of Connecticut Research Foundation.
\end{acknowledgments}


\bibliography{C:/Users/Public/bibliography/bibtex_masterfile_donotdelete}

\begin{thebibliography}{34}%
\makeatletter
\providecommand \@ifxundefined [1]{%
 \@ifx{#1\undefined}
}%
\providecommand \@ifnum [1]{%
 \ifnum #1\expandafter \@firstoftwo
 \else \expandafter \@secondoftwo
 \fi
}%
\providecommand \@ifx [1]{%
 \ifx #1\expandafter \@firstoftwo
 \else \expandafter \@secondoftwo
 \fi
}%
\providecommand \natexlab [1]{#1}%
\providecommand \enquote  [1]{``#1''}%
\providecommand \bibnamefont  [1]{#1}%
\providecommand \bibfnamefont [1]{#1}%
\providecommand \citenamefont [1]{#1}%
\providecommand \href@noop [0]{\@secondoftwo}%
\providecommand \href [0]{\begingroup \@sanitize@url \@href}%
\providecommand \@href[1]{\@@startlink{#1}\@@href}%
\providecommand \@@href[1]{\endgroup#1\@@endlink}%
\providecommand \@sanitize@url [0]{\catcode `\\12\catcode `\$12\catcode
  `\&12\catcode `\#12\catcode `\^12\catcode `\_12\catcode `\%12\relax}%
\providecommand \@@startlink[1]{}%
\providecommand \@@endlink[0]{}%
\providecommand \url  [0]{\begingroup\@sanitize@url \@url }%
\providecommand \@url [1]{\endgroup\@href {#1}{\urlprefix }}%
\providecommand \urlprefix  [0]{URL }%
\providecommand \Eprint [0]{\href }%
\providecommand \doibase [0]{http://dx.doi.org/}%
\providecommand \selectlanguage [0]{\@gobble}%
\providecommand \bibinfo  [0]{\@secondoftwo}%
\providecommand \bibfield  [0]{\@secondoftwo}%
\providecommand \translation [1]{[#1]}%
\providecommand \BibitemOpen [0]{}%
\providecommand \bibitemStop [0]{}%
\providecommand \bibitemNoStop [0]{.\EOS\space}%
\providecommand \EOS [0]{\spacefactor3000\relax}%
\providecommand \BibitemShut  [1]{\csname bibitem#1\endcsname}%
\let\auto@bib@innerbib\@empty
\bibitem [{\citenamefont {Lee}\ and\ \citenamefont {Mahan}(1965)}]{lee65}%
  \BibitemOpen
  \bibfield  {author} {\bibinfo {author} {\bibfnamefont {Y.~T.}\ \bibnamefont
  {Lee}}\ and\ \bibinfo {author} {\bibfnamefont {B.~H.}\ \bibnamefont
  {Mahan}},\ }\href {\doibase 10.1063/1.1703258} {\bibfield  {journal}
  {\bibinfo  {journal} {J. Chem. Phys.}\ }\textbf {\bibinfo {volume} {42}},\
  \bibinfo {pages} {2893} (\bibinfo {year} {1965})}\BibitemShut {NoStop}%
\bibitem [{\citenamefont {Klucharev}\ \emph {et~al.}(1980)\citenamefont
  {Klucharev}, \citenamefont {Lazarenko},\ and\ \citenamefont
  {Vujnovic}}]{klucharev80}%
  \BibitemOpen
  \bibfield  {author} {\bibinfo {author} {\bibfnamefont {A.~N.}\ \bibnamefont
  {Klucharev}}, \bibinfo {author} {\bibfnamefont {A.~V.}\ \bibnamefont
  {Lazarenko}}, \ and\ \bibinfo {author} {\bibfnamefont {V.}~\bibnamefont
  {Vujnovic}},\ }\href {http://stacks.iop.org/0022-3700/13/i=6/a=019}
  {\bibfield  {journal} {\bibinfo  {journal} {J. Phys. B}\ }\textbf {\bibinfo
  {volume} {13}},\ \bibinfo {pages} {1143} (\bibinfo {year}
  {1980})}\BibitemShut {NoStop}%
\bibitem [{\citenamefont {Wagner}\ and\ \citenamefont
  {Isenor}(1985)}]{wagner85}%
  \BibitemOpen
  \bibfield  {author} {\bibinfo {author} {\bibfnamefont {G.~S.}\ \bibnamefont
  {Wagner}}\ and\ \bibinfo {author} {\bibfnamefont {N.~R.}\ \bibnamefont
  {Isenor}},\ }\href {\doibase 10.1139/p85-160} {\bibfield  {journal} {\bibinfo
   {journal} {Can. J. Phys.}\ }\textbf {\bibinfo {volume} {63}},\ \bibinfo
  {pages} {976} (\bibinfo {year} {1985})}\BibitemShut {NoStop}%
\bibitem [{\citenamefont {Kappes}\ \emph {et~al.}(1985)\citenamefont {Kappes},
  \citenamefont {Schaer},\ and\ \citenamefont {Schumacher}}]{kappes85}%
  \BibitemOpen
  \bibfield  {author} {\bibinfo {author} {\bibfnamefont {M.~M.}\ \bibnamefont
  {Kappes}}, \bibinfo {author} {\bibfnamefont {M.}~\bibnamefont {Schaer}}, \
  and\ \bibinfo {author} {\bibfnamefont {E.}~\bibnamefont {Schumacher}},\
  }\href {\doibase 10.1021/j100254a038} {\bibfield  {journal} {\bibinfo
  {journal} {J. Phys. Chem.}\ }\textbf {\bibinfo {volume} {89}},\ \bibinfo
  {pages} {1499} (\bibinfo {year} {1985})}\BibitemShut {NoStop}%
\bibitem [{\citenamefont {Preuss}(1955)}]{preuss55}%
  \BibitemOpen
  \bibfield  {author} {\bibinfo {author} {\bibfnamefont {H.}~\bibnamefont
  {Preuss}},\ }\href@noop {} {\bibfield  {journal} {\bibinfo  {journal} {Z.
  Naturforsch. A.}\ }\textbf {\bibinfo {volume} {10}} (\bibinfo {year}
  {1955})}\BibitemShut {NoStop}%
\bibitem [{\citenamefont {Bellomonte}\ \emph {et~al.}(1974)\citenamefont
  {Bellomonte}, \citenamefont {Cavaliere},\ and\ \citenamefont
  {Ferrante}}]{bellomonte74}%
  \BibitemOpen
  \bibfield  {author} {\bibinfo {author} {\bibfnamefont {L.}~\bibnamefont
  {Bellomonte}}, \bibinfo {author} {\bibfnamefont {P.}~\bibnamefont
  {Cavaliere}}, \ and\ \bibinfo {author} {\bibfnamefont {G.}~\bibnamefont
  {Ferrante}},\ }\href {\doibase 10.1063/1.1682480} {\bibfield  {journal}
  {\bibinfo  {journal} {J. Chem. Phys.}\ }\textbf {\bibinfo {volume} {61}},\
  \bibinfo {pages} {3225} (\bibinfo {year} {1974})}\BibitemShut {NoStop}%
\bibitem [{\citenamefont {Patil}\ and\ \citenamefont {Tang}(2000)}]{patil00}%
  \BibitemOpen
  \bibfield  {author} {\bibinfo {author} {\bibfnamefont {S.~H.}\ \bibnamefont
  {Patil}}\ and\ \bibinfo {author} {\bibfnamefont {K.~T.}\ \bibnamefont
  {Tang}},\ }\href {\doibase 10.1063/1.481843} {\bibfield  {journal} {\bibinfo
  {journal} {J. Chem. Phys.}\ }\textbf {\bibinfo {volume} {113}},\ \bibinfo
  {pages} {676} (\bibinfo {year} {2000})}\BibitemShut {NoStop}%
\bibitem [{\citenamefont {Silberbach}\ \emph {et~al.}(1986)\citenamefont
  {Silberbach}, \citenamefont {Schwerdtfeger}, \citenamefont {Stoll},\ and\
  \citenamefont {Preuss}}]{silberbach86}%
  \BibitemOpen
  \bibfield  {author} {\bibinfo {author} {\bibfnamefont {H.}~\bibnamefont
  {Silberbach}}, \bibinfo {author} {\bibfnamefont {P.}~\bibnamefont
  {Schwerdtfeger}}, \bibinfo {author} {\bibfnamefont {H.}~\bibnamefont
  {Stoll}}, \ and\ \bibinfo {author} {\bibfnamefont {H.}~\bibnamefont
  {Preuss}},\ }\href {http://stacks.iop.org/0022-3700/19/i=5/a=011} {\bibfield
  {journal} {\bibinfo  {journal} {J. Phys. B}\ }\textbf {\bibinfo {volume}
  {19}},\ \bibinfo {pages} {501} (\bibinfo {year} {1986})}\BibitemShut
  {NoStop}%
\bibitem [{\citenamefont {von Szentp\'{a}ly}\ \emph {et~al.}(1982)\citenamefont
  {von Szentp\'{a}ly}, \citenamefont {Fuentealba}, \citenamefont {Preuss},\
  and\ \citenamefont {Stoll}}]{szentpaly82}%
  \BibitemOpen
  \bibfield  {author} {\bibinfo {author} {\bibfnamefont {L.}~\bibnamefont {von
  Szentp\'{a}ly}}, \bibinfo {author} {\bibfnamefont {P.}~\bibnamefont
  {Fuentealba}}, \bibinfo {author} {\bibfnamefont {H.}~\bibnamefont {Preuss}},
  \ and\ \bibinfo {author} {\bibfnamefont {H.}~\bibnamefont {Stoll}},\ }\href
  {\doibase 10.1016/0009-2614(82)83728-7} {\bibfield  {journal} {\bibinfo
  {journal} {Chem. Phys. Lett.}\ }\textbf {\bibinfo {volume} {93}},\ \bibinfo
  {pages} {555 } (\bibinfo {year} {1982})}\BibitemShut {NoStop}%
\bibitem [{\citenamefont {Krauss}\ and\ \citenamefont
  {Stevens}(1990)}]{krauss90}%
  \BibitemOpen
  \bibfield  {author} {\bibinfo {author} {\bibfnamefont {M.}~\bibnamefont
  {Krauss}}\ and\ \bibinfo {author} {\bibfnamefont {W.~J.}\ \bibnamefont
  {Stevens}},\ }\href {\doibase 10.1063/1.458756} {\bibfield  {journal}
  {\bibinfo  {journal} {The J. Chem. Phys.}\ }\textbf {\bibinfo {volume}
  {93}},\ \bibinfo {pages} {4236} (\bibinfo {year} {1990})}\BibitemShut
  {NoStop}%
\bibitem [{\citenamefont {Aymar}\ \emph {et~al.}(2003)\citenamefont {Aymar},
  \citenamefont {Azizi},\ and\ \citenamefont {Dulieu}}]{aymar03}%
  \BibitemOpen
  \bibfield  {author} {\bibinfo {author} {\bibfnamefont {M.}~\bibnamefont
  {Aymar}}, \bibinfo {author} {\bibfnamefont {S.}~\bibnamefont {Azizi}}, \ and\
  \bibinfo {author} {\bibfnamefont {O.}~\bibnamefont {Dulieu}},\ }\href
  {http://stacks.iop.org/0953-4075/36/i=24/a=004} {\bibfield  {journal}
  {\bibinfo  {journal} {J. Phys. B}\ }\textbf {\bibinfo {volume} {36}},\
  \bibinfo {pages} {4799} (\bibinfo {year} {2003})}\BibitemShut {NoStop}%
\bibitem [{\citenamefont {Jraij}\ \emph {et~al.}(2003)\citenamefont {Jraij},
  \citenamefont {Allouche}, \citenamefont {Korek},\ and\ \citenamefont
  {Aubert-Fr\'{e}con}}]{jraij03}%
  \BibitemOpen
  \bibfield  {author} {\bibinfo {author} {\bibfnamefont {A.}~\bibnamefont
  {Jraij}}, \bibinfo {author} {\bibfnamefont {A.}~\bibnamefont {Allouche}},
  \bibinfo {author} {\bibfnamefont {M.}~\bibnamefont {Korek}}, \ and\ \bibinfo
  {author} {\bibfnamefont {M.}~\bibnamefont {Aubert-Fr\'{e}con}},\ }\href
  {\doibase 10.1016/S0301-0104(03)00060-0} {\bibfield  {journal} {\bibinfo
  {journal} {Chem. Phys.}\ }\textbf {\bibinfo {volume} {290}},\ \bibinfo
  {pages} {129 } (\bibinfo {year} {2003})}\BibitemShut {NoStop}%
\bibitem [{\citenamefont {Roche}\ and\ \citenamefont {Jungen}(1993)}]{roche93}%
  \BibitemOpen
  \bibfield  {author} {\bibinfo {author} {\bibfnamefont {A.~L.}\ \bibnamefont
  {Roche}}\ and\ \bibinfo {author} {\bibfnamefont {C.}~\bibnamefont {Jungen}},\
  }\href {\doibase 10.1063/1.464040} {\bibfield  {journal} {\bibinfo  {journal}
  {J. Chem. Phys.}\ }\textbf {\bibinfo {volume} {98}},\ \bibinfo {pages} {3637}
  (\bibinfo {year} {1993})}\BibitemShut {NoStop}%
\bibitem [{\citenamefont {Chang}\ \emph {et~al.}(1999)\citenamefont {Chang},
  \citenamefont {Li}, \citenamefont {Zhang}, \citenamefont {Tsai},
  \citenamefont {Bahns},\ and\ \citenamefont {Stwalley}}]{chang99}%
  \BibitemOpen
  \bibfield  {author} {\bibinfo {author} {\bibfnamefont {E.~S.}\ \bibnamefont
  {Chang}}, \bibinfo {author} {\bibfnamefont {J.}~\bibnamefont {Li}}, \bibinfo
  {author} {\bibfnamefont {J.}~\bibnamefont {Zhang}}, \bibinfo {author}
  {\bibfnamefont {C.-C.}\ \bibnamefont {Tsai}}, \bibinfo {author}
  {\bibfnamefont {J.}~\bibnamefont {Bahns}}, \ and\ \bibinfo {author}
  {\bibfnamefont {W.~C.}\ \bibnamefont {Stwalley}},\ }\href {\doibase
  10.1063/1.479929} {\bibfield  {journal} {\bibinfo  {journal} {J. Chem.
  Phys.}\ }\textbf {\bibinfo {volume} {111}},\ \bibinfo {pages} {6247}
  (\bibinfo {year} {1999})}\BibitemShut {NoStop}%
\bibitem [{\citenamefont {Broyer}\ \emph {et~al.}(1983)\citenamefont {Broyer},
  \citenamefont {Chevaleyre}, \citenamefont {Delacretaz}, \citenamefont
  {Martin},\ and\ \citenamefont {W{\"{o}}ste}}]{broyer83}%
  \BibitemOpen
  \bibfield  {author} {\bibinfo {author} {\bibfnamefont {M.}~\bibnamefont
  {Broyer}}, \bibinfo {author} {\bibfnamefont {J.}~\bibnamefont {Chevaleyre}},
  \bibinfo {author} {\bibfnamefont {G.}~\bibnamefont {Delacretaz}}, \bibinfo
  {author} {\bibfnamefont {S.}~\bibnamefont {Martin}}, \ and\ \bibinfo {author}
  {\bibfnamefont {L.}~\bibnamefont {W{\"{o}}ste}},\ }\href {\doibase
  10.1016/0009-2614(83)87525-3} {\bibfield  {journal} {\bibinfo  {journal}
  {Chem. Phys. Lett.}\ }\textbf {\bibinfo {volume} {99}},\ \bibinfo {pages}
  {206 } (\bibinfo {year} {1983})}\BibitemShut {NoStop}%
\bibitem [{\citenamefont {Kim}\ and\ \citenamefont {Yoshihara}(1993)}]{kim93}%
  \BibitemOpen
  \bibfield  {author} {\bibinfo {author} {\bibfnamefont {B.}~\bibnamefont
  {Kim}}\ and\ \bibinfo {author} {\bibfnamefont {K.}~\bibnamefont
  {Yoshihara}},\ }\href {\doibase 10.1016/0009-2614(93)90067-B} {\bibfield
  {journal} {\bibinfo  {journal} {Chem. Phys. Lett.}\ }\textbf {\bibinfo
  {volume} {202}},\ \bibinfo {pages} {437 } (\bibinfo {year}
  {1993})}\BibitemShut {NoStop}%
\bibitem [{\citenamefont {Stwalley}\ and\ \citenamefont
  {Bahns}(1993)}]{stwalley93}%
  \BibitemOpen
  \bibfield  {author} {\bibinfo {author} {\bibfnamefont {W.~C.}\ \bibnamefont
  {Stwalley}}\ and\ \bibinfo {author} {\bibfnamefont {J.~T.}\ \bibnamefont
  {Bahns}},\ }\href {\doibase 10.1017/S0263034600007047} {\bibfield  {journal}
  {\bibinfo  {journal} {Laser and Particle Beams}\ }\textbf {\bibinfo {volume}
  {11}},\ \bibinfo {pages} {185} (\bibinfo {year} {1993})}\BibitemShut
  {NoStop}%
\bibitem [{\citenamefont {Strauss}\ \emph {et~al.}(2010)\citenamefont
  {Strauss}, \citenamefont {Takekoshi}, \citenamefont {Lang}, \citenamefont
  {Winkler}, \citenamefont {Grimm}, \citenamefont {Hecker~Denschlag},\ and\
  \citenamefont {Tiemann}}]{strauss10}%
  \BibitemOpen
  \bibfield  {author} {\bibinfo {author} {\bibfnamefont {C.}~\bibnamefont
  {Strauss}}, \bibinfo {author} {\bibfnamefont {T.}~\bibnamefont {Takekoshi}},
  \bibinfo {author} {\bibfnamefont {F.}~\bibnamefont {Lang}}, \bibinfo {author}
  {\bibfnamefont {K.}~\bibnamefont {Winkler}}, \bibinfo {author} {\bibfnamefont
  {R.}~\bibnamefont {Grimm}}, \bibinfo {author} {\bibfnamefont
  {J.}~\bibnamefont {Hecker~Denschlag}}, \ and\ \bibinfo {author}
  {\bibfnamefont {E.}~\bibnamefont {Tiemann}},\ }\href {\doibase
  10.1103/PhysRevA.82.052514} {\bibfield  {journal} {\bibinfo  {journal} {Phys.
  Rev. A}\ }\textbf {\bibinfo {volume} {82}},\ \bibinfo {pages} {052514}
  (\bibinfo {year} {2010})}\BibitemShut {NoStop}%
\bibitem [{\citenamefont {Lefebvre-Brion}\ and\ \citenamefont
  {Field}(2004)}]{lefevbre-brion04}%
  \BibitemOpen
  \bibfield  {author} {\bibinfo {author} {\bibfnamefont {H.}~\bibnamefont
  {Lefebvre-Brion}}\ and\ \bibinfo {author} {\bibfnamefont {R.~W.}\
  \bibnamefont {Field}},\ }\href {http://amazon.com/o/ASIN/0124414567/} {\emph
  {\bibinfo {title} {The Spectra and Dynamics of Diatomic Molecules: Revised
  and Enlarged Edition}}}\ (\bibinfo  {publisher} {Academic Press, Amsterdam},\
  \bibinfo {year} {2004})\BibitemShut {NoStop}%
\bibitem [{\citenamefont {Le{$\:$}Roy}(2007)}]{level8}%
  \BibitemOpen
  \bibfield  {author} {\bibinfo {author} {\bibfnamefont {R.~J.}\ \bibnamefont
  {Le{$\:$}Roy}},\ }\href {http://scienide2.uwaterloo.ca/~rleroy/level/}
  {\bibfield  {journal} {\bibinfo  {journal} {\emph{LEVEL8.0}: A Computer
  Program for Solving the Radial Schr{\"{o}}dinger Equation for Bound and
  Quasibound Levels, University of Waterloo Chemical Physics Research Report
  CP-663}\ } (\bibinfo {year} {2007})}\BibitemShut {NoStop}%
\bibitem [{\citenamefont {Amiot}(1990)}]{amiot93}%
  \BibitemOpen
  \bibfield  {author} {\bibinfo {author} {\bibfnamefont {C.}~\bibnamefont
  {Amiot}},\ }\href {\doibase 10.1063/1.459246} {\bibfield  {journal} {\bibinfo
   {journal} {J Chem. Phys.}\ }\textbf {\bibinfo {volume} {93}},\ \bibinfo
  {pages} {8591} (\bibinfo {year} {1990})}\BibitemShut {NoStop}%
\bibitem [{\citenamefont {Tsai}\ \emph {et~al.}(1997)\citenamefont {Tsai},
  \citenamefont {Freeland}, \citenamefont {Vogels}, \citenamefont {Boesten},
  \citenamefont {Verhaar},\ and\ \citenamefont {Heinzen}}]{tsai97}%
  \BibitemOpen
  \bibfield  {author} {\bibinfo {author} {\bibfnamefont {C.~C.}\ \bibnamefont
  {Tsai}}, \bibinfo {author} {\bibfnamefont {R.~S.}\ \bibnamefont {Freeland}},
  \bibinfo {author} {\bibfnamefont {J.~M.}\ \bibnamefont {Vogels}}, \bibinfo
  {author} {\bibfnamefont {H.~M. J.~M.}\ \bibnamefont {Boesten}}, \bibinfo
  {author} {\bibfnamefont {B.~J.}\ \bibnamefont {Verhaar}}, \ and\ \bibinfo
  {author} {\bibfnamefont {D.~J.}\ \bibnamefont {Heinzen}},\ }\href {\doibase
  10.1103/PhysRevLett.79.1245} {\bibfield  {journal} {\bibinfo  {journal}
  {Phys. Rev. Lett.}\ }\textbf {\bibinfo {volume} {79}},\ \bibinfo {pages}
  {1245} (\bibinfo {year} {1997})}\BibitemShut {NoStop}%
\bibitem [{\citenamefont {Seto}\ \emph {et~al.}(2000)\citenamefont {Seto},
  \citenamefont {LeRoy}, \citenamefont {Verg\`{e}s},\ and\ \citenamefont
  {Amiot}}]{seto00}%
  \BibitemOpen
  \bibfield  {author} {\bibinfo {author} {\bibfnamefont {J.~Y.}\ \bibnamefont
  {Seto}}, \bibinfo {author} {\bibfnamefont {R.~J.}\ \bibnamefont {LeRoy}},
  \bibinfo {author} {\bibfnamefont {J.}~\bibnamefont {Verg\`{e}s}}, \ and\
  \bibinfo {author} {\bibfnamefont {C.}~\bibnamefont {Amiot}},\ }\href
  {\doibase 10.1063/1.1286979} {\bibfield  {journal} {\bibinfo  {journal} {J.
  Chem. Phys.}\ }\textbf {\bibinfo {volume} {113}},\ \bibinfo {pages} {3067}
  (\bibinfo {year} {2000})}\BibitemShut {NoStop}%
\bibitem [{\citenamefont {Beser}\ \emph {et~al.}(2009)\citenamefont {Beser},
  \citenamefont {Sovkov}, \citenamefont {Bai}, \citenamefont {Ahmed},
  \citenamefont {Tsai}, \citenamefont {Xie}, \citenamefont {Li}, \citenamefont
  {Ivanov},\ and\ \citenamefont {Lyyra}}]{beser09}%
  \BibitemOpen
  \bibfield  {author} {\bibinfo {author} {\bibfnamefont {B.}~\bibnamefont
  {Beser}}, \bibinfo {author} {\bibfnamefont {V.~B.}\ \bibnamefont {Sovkov}},
  \bibinfo {author} {\bibfnamefont {J.}~\bibnamefont {Bai}}, \bibinfo {author}
  {\bibfnamefont {E.~H.}\ \bibnamefont {Ahmed}}, \bibinfo {author}
  {\bibfnamefont {C.~C.}\ \bibnamefont {Tsai}}, \bibinfo {author}
  {\bibfnamefont {F.}~\bibnamefont {Xie}}, \bibinfo {author} {\bibfnamefont
  {L.}~\bibnamefont {Li}}, \bibinfo {author} {\bibfnamefont {V.~S.}\
  \bibnamefont {Ivanov}}, \ and\ \bibinfo {author} {\bibfnamefont {A.~M.}\
  \bibnamefont {Lyyra}},\ }\href {\doibase 10.1063/1.3194290} {\bibfield
  {journal} {\bibinfo  {journal} {J. Chem. Phys.}\ }\textbf {\bibinfo {volume}
  {131}},\ \bibinfo {eid} {094505} (\bibinfo {year} {2009})}\BibitemShut
  {NoStop}%
\bibitem [{\citenamefont {Bellos}\ \emph {et~al.}(2012)\citenamefont {Bellos},
  \citenamefont {Carollo}, \citenamefont {Rahmlow}, \citenamefont {Banerjee},
  \citenamefont {Eyler}, \citenamefont {Gould},\ and\ \citenamefont
  {Stwalley}}]{bellos12}%
  \BibitemOpen
  \bibfield  {author} {\bibinfo {author} {\bibfnamefont {M.~A.}\ \bibnamefont
  {Bellos}}, \bibinfo {author} {\bibfnamefont {R.}~\bibnamefont {Carollo}},
  \bibinfo {author} {\bibfnamefont {D.}~\bibnamefont {Rahmlow}}, \bibinfo
  {author} {\bibfnamefont {J.}~\bibnamefont {Banerjee}}, \bibinfo {author}
  {\bibfnamefont {E.~E.}\ \bibnamefont {Eyler}}, \bibinfo {author}
  {\bibfnamefont {P.~L.}\ \bibnamefont {Gould}}, \ and\ \bibinfo {author}
  {\bibfnamefont {W.~C.}\ \bibnamefont {Stwalley}},\ }\href {\doibase
  10.1103/PhysRevA.86.033407} {\bibfield  {journal} {\bibinfo  {journal} {Phys.
  Rev. A}\ }\textbf {\bibinfo {volume} {86}},\ \bibinfo {pages} {033407}
  (\bibinfo {year} {2012})}\BibitemShut {NoStop}%
\bibitem [{\citenamefont {Lozeille}\ \emph {et~al.}(2006)\citenamefont
  {Lozeille}, \citenamefont {Fioretti}, \citenamefont {Gabbanini},
  \citenamefont {Huang}, \citenamefont {Pechkis}, \citenamefont {Wang},
  \citenamefont {Gould}, \citenamefont {Eyler}, \citenamefont {Stwalley},
  \citenamefont {Aymar},\ and\ \citenamefont {Dulieu}}]{lozeille06}%
  \BibitemOpen
  \bibfield  {author} {\bibinfo {author} {\bibfnamefont {J.}~\bibnamefont
  {Lozeille}}, \bibinfo {author} {\bibfnamefont {A.}~\bibnamefont {Fioretti}},
  \bibinfo {author} {\bibfnamefont {C.}~\bibnamefont {Gabbanini}}, \bibinfo
  {author} {\bibfnamefont {Y.}~\bibnamefont {Huang}}, \bibinfo {author}
  {\bibfnamefont {H.~K.}\ \bibnamefont {Pechkis}}, \bibinfo {author}
  {\bibfnamefont {D.}~\bibnamefont {Wang}}, \bibinfo {author} {\bibfnamefont
  {P.~L.}\ \bibnamefont {Gould}}, \bibinfo {author} {\bibfnamefont {E.~E.}\
  \bibnamefont {Eyler}}, \bibinfo {author} {\bibfnamefont {W.~C.}\ \bibnamefont
  {Stwalley}}, \bibinfo {author} {\bibfnamefont {M.}~\bibnamefont {Aymar}}, \
  and\ \bibinfo {author} {\bibfnamefont {O.}~\bibnamefont {Dulieu}},\ }\href
  {http://dx.doi.org/10.1140/epjd/e2006-00084-4} {\bibfield  {journal}
  {\bibinfo  {journal} {Eur. Phys. J. D}\ }\textbf {\bibinfo {volume} {39}},\
  \bibinfo {pages} {261} (\bibinfo {year} {2006})}\BibitemShut {NoStop}%
\bibitem [{\citenamefont {Huang}\ \emph {et~al.}(2006)\citenamefont {Huang},
  \citenamefont {Qi}, \citenamefont {Pechkis}, \citenamefont {Wang},
  \citenamefont {Eyler}, \citenamefont {Gould},\ and\ \citenamefont
  {Stwalley}}]{huang06}%
  \BibitemOpen
  \bibfield  {author} {\bibinfo {author} {\bibfnamefont {Y.}~\bibnamefont
  {Huang}}, \bibinfo {author} {\bibfnamefont {J.}~\bibnamefont {Qi}}, \bibinfo
  {author} {\bibfnamefont {H.~K.}\ \bibnamefont {Pechkis}}, \bibinfo {author}
  {\bibfnamefont {D.}~\bibnamefont {Wang}}, \bibinfo {author} {\bibfnamefont
  {E.~E.}\ \bibnamefont {Eyler}}, \bibinfo {author} {\bibfnamefont {P.~L.}\
  \bibnamefont {Gould}}, \ and\ \bibinfo {author} {\bibfnamefont {W.~C.}\
  \bibnamefont {Stwalley}},\ }\href
  {http://stacks.iop.org/0953-4075/39/i=19/a=S04} {\bibfield  {journal}
  {\bibinfo  {journal} {Journal of Physics B: Atomic, Molecular and Optical
  Physics}\ }\textbf {\bibinfo {volume} {39}},\ \bibinfo {pages} {S857}
  (\bibinfo {year} {2006})}\BibitemShut {NoStop}%
\bibitem [{\citenamefont {Gallagher}(1994)}]{gallagher94}%
  \BibitemOpen
  \bibfield  {author} {\bibinfo {author} {\bibfnamefont {T.~F.}\ \bibnamefont
  {Gallagher}},\ }\href@noop {} {\emph {\bibinfo {title} {Rydberg Atoms}}}\
  (\bibinfo  {publisher} {Cambridge University Press, Cambridge},\ \bibinfo
  {year} {1994})\BibitemShut {NoStop}%
\bibitem [{\citenamefont {Machholm}\ \emph {et~al.}(1994)\citenamefont
  {Machholm}, \citenamefont {Giusti-Suzor},\ and\ \citenamefont
  {Mies}}]{machholm94}%
  \BibitemOpen
  \bibfield  {author} {\bibinfo {author} {\bibfnamefont {M.}~\bibnamefont
  {Machholm}}, \bibinfo {author} {\bibfnamefont {A.}~\bibnamefont
  {Giusti-Suzor}}, \ and\ \bibinfo {author} {\bibfnamefont {F.~H.}\
  \bibnamefont {Mies}},\ }\href {\doibase 10.1103/PhysRevA.50.5025} {\bibfield
  {journal} {\bibinfo  {journal} {Phys. Rev. A}\ }\textbf {\bibinfo {volume}
  {50}},\ \bibinfo {pages} {5025} (\bibinfo {year} {1994})}\BibitemShut
  {NoStop}%
\bibitem [{\citenamefont {Amelink}\ \emph {et~al.}(2000)\citenamefont
  {Amelink}, \citenamefont {Jones}, \citenamefont {Lett}, \citenamefont
  {Straten},\ and\ \citenamefont {Heideman}}]{amelink00}%
  \BibitemOpen
  \bibfield  {author} {\bibinfo {author} {\bibfnamefont {A.}~\bibnamefont
  {Amelink}}, \bibinfo {author} {\bibfnamefont {K.~M.}\ \bibnamefont {Jones}},
  \bibinfo {author} {\bibfnamefont {P.~D.}\ \bibnamefont {Lett}}, \bibinfo
  {author} {\bibfnamefont {P.~v.~d.}\ \bibnamefont {Straten}}, \ and\ \bibinfo
  {author} {\bibfnamefont {H.~G.~M.}\ \bibnamefont {Heideman}},\ }\href
  {\doibase 10.1103/PhysRevA.61.042707} {\bibfield  {journal} {\bibinfo
  {journal} {Phys. Rev. A}\ }\textbf {\bibinfo {volume} {61}},\ \bibinfo
  {pages} {042707} (\bibinfo {year} {2000})}\BibitemShut {NoStop}%
\bibitem [{\citenamefont {Trachy}\ \emph {et~al.}(2007)\citenamefont {Trachy},
  \citenamefont {Veshapidze}, \citenamefont {Shah}, \citenamefont {Jang},\ and\
  \citenamefont {DePaola}}]{trachy07}%
  \BibitemOpen
  \bibfield  {author} {\bibinfo {author} {\bibfnamefont {M.~L.}\ \bibnamefont
  {Trachy}}, \bibinfo {author} {\bibfnamefont {G.}~\bibnamefont {Veshapidze}},
  \bibinfo {author} {\bibfnamefont {M.~H.}\ \bibnamefont {Shah}}, \bibinfo
  {author} {\bibfnamefont {H.~U.}\ \bibnamefont {Jang}}, \ and\ \bibinfo
  {author} {\bibfnamefont {B.~D.}\ \bibnamefont {DePaola}},\ }\href {\doibase
  10.1103/PhysRevLett.99.043003} {\bibfield  {journal} {\bibinfo  {journal}
  {Phys. Rev. Lett.}\ }\textbf {\bibinfo {volume} {99}},\ \bibinfo {pages}
  {043003} (\bibinfo {year} {2007})}\BibitemShut {NoStop}%
\bibitem [{\citenamefont {Allouche}\ and\ \citenamefont
  {Aubert-Fr\'{e}con}(2012)}]{allouche12}%
  \BibitemOpen
  \bibfield  {author} {\bibinfo {author} {\bibfnamefont {A.-R.}\ \bibnamefont
  {Allouche}}\ and\ \bibinfo {author} {\bibfnamefont {M.}~\bibnamefont
  {Aubert-Fr\'{e}con}},\ }\href {\doibase 10.1063/1.3694014} {\bibfield
  {journal} {\bibinfo  {journal} {J. Chem. Phys.}\ }\textbf {\bibinfo {volume}
  {136}},\ \bibinfo {eid} {114302} (\bibinfo {year} {2012})}\BibitemShut
  {NoStop}%
\bibitem [{\citenamefont {Sanguinetti}\ \emph {et~al.}(2009)\citenamefont
  {Sanguinetti}, \citenamefont {Majeed}, \citenamefont {Jones},\ and\
  \citenamefont {Varcoe}}]{sanguinetti09}%
  \BibitemOpen
  \bibfield  {author} {\bibinfo {author} {\bibfnamefont {B.}~\bibnamefont
  {Sanguinetti}}, \bibinfo {author} {\bibfnamefont {H.~O.}\ \bibnamefont
  {Majeed}}, \bibinfo {author} {\bibfnamefont {M.~L.}\ \bibnamefont {Jones}}, \
  and\ \bibinfo {author} {\bibfnamefont {B.~T.~H.}\ \bibnamefont {Varcoe}},\
  }\href {http://stacks.iop.org/0953-4075/42/i=16/a=165004} {\bibfield
  {journal} {\bibinfo  {journal} {J. Phys. B}\ }\textbf {\bibinfo {volume}
  {42}},\ \bibinfo {pages} {165004} (\bibinfo {year} {2009})}\BibitemShut
  {NoStop}%
\bibitem [{\citenamefont {Bellos}\ \emph {et~al.}(2011)\citenamefont {Bellos},
  \citenamefont {Rahmlow}, \citenamefont {Carollo}, \citenamefont {Banerjee},
  \citenamefont {Dulieu}, \citenamefont {Gerdes}, \citenamefont {Eyler},
  \citenamefont {Gould},\ and\ \citenamefont {Stwalley}}]{bellos11}%
  \BibitemOpen
  \bibfield  {author} {\bibinfo {author} {\bibfnamefont {M.~A.}\ \bibnamefont
  {Bellos}}, \bibinfo {author} {\bibfnamefont {D.}~\bibnamefont {Rahmlow}},
  \bibinfo {author} {\bibfnamefont {R.}~\bibnamefont {Carollo}}, \bibinfo
  {author} {\bibfnamefont {J.}~\bibnamefont {Banerjee}}, \bibinfo {author}
  {\bibfnamefont {O.}~\bibnamefont {Dulieu}}, \bibinfo {author} {\bibfnamefont
  {A.}~\bibnamefont {Gerdes}}, \bibinfo {author} {\bibfnamefont {E.~E.}\
  \bibnamefont {Eyler}}, \bibinfo {author} {\bibfnamefont {P.~L.}\ \bibnamefont
  {Gould}}, \ and\ \bibinfo {author} {\bibfnamefont {W.~C.}\ \bibnamefont
  {Stwalley}},\ }\href {\doibase 10.1039/C1CP21383K} {\bibfield  {journal}
  {\bibinfo  {journal} {Phys. Chem. Chem. Phys.}\ }\textbf {\bibinfo {volume}
  {13}},\ \bibinfo {pages} {18880} (\bibinfo {year} {2011})}\BibitemShut
  {NoStop}%
\end{thebibliography}%

\end{document}